\documentclass[useAMS,usenatbib]{mn2e}
\usepackage{epsfig}
\usepackage{amsmath}

\newcommand{\be}{\begin{equation}}
\newcommand{\beq}{\begin{equation}}
\newcommand{\ba}{\begin{eqnarray}}
\newcommand{\ee}{\end{equation}}
\newcommand{\eeq}{\end{equation}}
\newcommand{\ea}{\end{eqnarray}}

\newcommand{\apj}{ApJ}
\newcommand{\apjl}{ApJL}
\newcommand{\mnras}{MNRAS}

\newcommand{\apjs}{ApJS}

\def\lsim{~\rlap{$<$}{\lower 1.0ex\hbox{$\sim$}}}

\def\gsim{~\rlap{$>$}{\lower 1.0ex\hbox{$\sim$}}}

\title[The HOD of Neutral Hydrogen]{The Halo Occupation Distribution of HI from 21cm Intensity Mapping at Moderate Redshifts}

\author[Wyithe \& Brown]{J. Stuart B. Wyithe$^1$, Michael J. I. Brown$^2$\\$^1$School of Physics, University of Melbourne, Parkville, Victoria,
Australia\\$^2$ School of Physics, Monash University, Clayton, Victoria 3800, Australia\\
Email: swyithe@unimelb.edu.au}

\begin{document}


\maketitle

\label{firstpage}
\begin{abstract}

  \noindent The spatial clustering properties of HI galaxies can be
  studied using the formalism of the halo occupation distribution
  (HOD). The resulting parameter constraints describe properties like
  gas richness verses environment. Unfortunately, clustering studies
  based on individual HI galaxies will be restricted to the local
  Universe for the foreseeable future, even with the deepest HI
  surveys. Here we discuss how clustering studies of the HI HOD could
  be extended to moderate redshift, through observations of
  fluctuations in the combined 21cm intensity of unresolved
  galaxies. In particular we make an analytic estimate for the
  clustering of HI in the HOD. Our joint goals are to estimate $i)$
  the amplitude of the signal, and $ii)$ the sensitivity of telescopes
  like the Australian SKA Pathfinder to HOD parameters.  We find that
  the power spectrum of redshifted 21cm intensity could be used to
  study the distribution of HI within dark matter halos at $z\ga0.5$
  where individual galaxies cannot be detected. In addition to the HOD
  of HI, the amplitude of the 21cm power spectrum would also yield
  estimates of the cosmic HI content ($\Omega_{\rm HI}$) at epochs
  between the local Universe, and redshifts probed by damped
  Ly$\alpha$ absorbers.

\end{abstract}


\section{Introduction}

The cosmic star-formation rate has declined by more than an order of
magnitude in the 8 billion years since $z\sim1$ (Lilly et al. 1996, Madau et
al. 1996).  Why this decline has taken place, and what drove it are
two of the most important unanswered questions in our current
understanding of galaxy formation and evolution.  In cold dark matter
cosmologies, gas cools and collapses to form stars within
gravitationally bound "halos" of dark matter. These galaxies can then
grow via continued star formation or via mergers with other
galaxies. As a result the decline in star formation at $z\la1$ is presumably
accompanied by a decrease in the amount of cold gas within halos.  

One of the issues that will need to be addressed in order to understand the evolution in star formation rate is the role of environment.  
As galaxies of a given baryonic mass can only reside within dark
matter halos above a particular dark matter mass, galaxies are biased
tracers of the overall dark matter distribution. The clustering of
dark matter halos is a known function of their mass (e.g., Sheth, Mo
\& Tormen 2001), and consequently the large-scale clustering of
galaxies provides an estimate of the typical halo mass in which that
galaxy population resides. On smaller scales, multiple galaxies can
reside within a single ($\la1$ Mpc radius) dark matter halo, so that
the number of galaxy pairs with small spatial separations is a strong
function of the number of galaxies per halo. One can thus constrain
the number of galaxies per halo as a function of halo mass by
measuring both the small ($\la1$ Mpc) and large ($\ga1$ Mpc) scale
clustering of galaxies (e.g.  Peacock \& Smith 2000, Zheng 2005).
The clustering of galaxy samples selected to lie within
different stages of galaxy formation based on their stellar and cold
gas content therefore has the potential to play a central role in our understanding of the star formation
history.

In recent years large galaxy redshift surveys such as SDSS and the
2dFGRS have enabled detailed studies of the clustering of more than
100,000 optically selected galaxies in the nearby universe.  By using
clustering to understand how galaxies populate dark matter halos, key
insights may be obtained into how galaxies grow over cosmic time. The
way in which stellar mass populates dark matter halos has been
determined through studies of clustering for optically selected galaxy
samples.  A popular formalism for modeling clustering on small to
large scales is termed the halo occupation distribution (HOD; e.g. Peacock \& Smith~2000; Seljak 2000; Scoccimarro et
al.~2001; Berlind \& Weinberg~2002; Zheng~2004). The HOD
includes contributions to galaxy clustering from pairs of galaxies in
distinct halos which describes the clustering in the large scale
limit, and from pairs of galaxies within a single halo which describes
clustering in the small scale limit. The latter contribution requires
a parametrisation to relate the number and spatial distribution of
galaxies within a dark matter halo of a particular mass. It is by
constraining this parameterisation that observed clustering can be
used to understand how galaxies are distributed.
 
By comparison with the massive optical redshift surveys, the largest
survey of HI selected galaxies contains only $\sim5000$ sources,
obtained as part of the HIPASS survey, a blind HI survey of the
southern sky (Barnes et al.~2001).  Meyer et al.~(2007) have studied
the clustering of these HI galaxies.  Their analysis reached the
conclusion of weak clustering of HI galaxies based on parametric
estimates of correlation length, but did not study the clustering in
terms of the host dark matter halo masses of the HIPASS sample.
Wyithe, Brown, Zwaan \& Meyer~(2009) analysed the clustering
properties of HI selected galaxies from the HIPASS survey using the
formalism of the halo occupation distribution. They found that the
real-space clustering amplitude for HIPASS galaxies is significant on
scales below the virial radius associated with the halo mass required
to reproduce the clustering amplitude on large scales, indicating that
single halo pairs are contributing a 1-halo term. However the
resulting parameter constraints show that satellite galaxies make up
only $\sim10\%$ of the HIPASS sample. HI satellite galaxies are
therefore less significant in number and in terms of their
contribution to clustering statistics than are satellites in optically
selected galaxy redshift surveys. 

These results from HOD modeling of HI galaxy
clustering therefore quantify the extent to which environment governs the HI
content of galaxies in the local Universe and confirms previous
evidence that HI galaxies are relatively rare in overdense
environments (Waugh et al.~2002; Cortes et al.~2008). Wyithe et
al.~(2009) found a minimum halo mass for HIPASS galaxies at the peak
of the redshift distribution of $M\sim10^{11}$M$_\odot$ (throughout
this paper we refer to the halo mass as $M$ and the HI mass as $M_{\rm
  HI}$), and showed that less than 10\% of baryons in HIPASS galaxies
are in the form of HI. Their analysis also revealed that the
fingers-of-god in the redshift space correlation function are
sensitive to the typical halo mass in which satellite galaxies reside,
and indicated that the HI rich satellites required to produce the
measured 1-halo term must be preferentially in group rather than
cluster mass halos.

As described above, the clustering of HI galaxies can be studied at
$z=0$ using HIPASS.  However in the future with the advent of the
Square Kilometer Array (SKA) and its pathfinders the volume and
redshift range over which clustering of HI galaxies can be studied
will greatly increase. On the other hand, these studies will still be
limited to moderate redshifts of $z\la0.5$ owing to the sensitivity
required for detection of even the most massive HI galaxies. For
example, the Australian SKA pathfinder (ASKAP) will detect the most
massive HI galaxies only out to $z\sim0.7$ in the deepest integrations
(Johnston et al.~2007). At higher redshifts, we argue that progress on the clustering of HI
galaxies may be made by measurement of fluctuations in the combined
surface-brightness of unresolved HI galaxies (Wyithe \& Loeb~2009;
Chang et al.~2008; Wyithe~2008).
A survey of 21cm intensity fluctuations at redshifts beyond those
where individual galaxies can be detected would therefore measure the
modulation of the cumulative 21cm emission from a large number of
galaxies. The detectability of the 21~cm PS after reionization was
discussed by Khandai et al.~(2009). These authors used an N-body and 
simulation to predict the statistical signal of 21~cm fluctuations in
the post-reionization IGM, and estimated its detectability. 
Khandai et al.~(2009) find that a
combination of these arrays offer good prospects for detecting the
21~cm PS over a range of redshifts in the post reionization era.
Importantly, a statistical detection of 21~cm fluctuations due to
discrete, unresolved clumps of neutral gas has already been made
(Pen et al.~2008) through cross-correlation of the HIPASS
(Barnes et al.~2001) 21~cm observations of the local universe with
galaxies in the 6 degree field galaxy redshift survey
(Jones et al.~2005). This detection represents an important step towards
using 21~cm surface brightness fluctuations to probe the neutral gas
distribution in the IGM.

The majority of the discussion in the literature concerning 21cm
fluctuations in the low-redshift Universe has centered around their
utility for cosmological constraint (Wyithe \& Loeb~2009; Chang et
al.~2008; Loeb \& Wyithe~2009; Bharadwaj, Sethi \& Saini~2009). In
this paper we concentrate on the possibility of studying the
distribution of HI within dark matter halos, on scales accessible to
traditional configurations for radio interferometers (which do not
include the very short baselines required to study the 21cm
fluctuations in the large scale, linear regime).  
On these scales recent simulations suggest that the smoothed HI density field is highly biased owing to non-linear gravitational clustering (Bagla \& Khandai~2009). Following from this prediction we
discuss the possibility of studying the occupation of dark matter
halos by HI at high redshift via 21cm intensity mapping. As a concrete
example we consider the potential of ASKAP with respect to
constraining the HI HOD. We concentrate on $z\sim0.7$, as this is the
redshift at which ASKAP no longer has the sensitivity to study
individual galaxies. Our goal is not to provide a detailed method for
extracting detailed HOD parameters from an observed power spectrum of
redshifted 21cm fluctuations. This would require calibration against
N-body simulations, which is premature at this time. Rather, we
present an analytic model for the 21cm power spectrum in the HOD, and
investigate which of its properties could be constrained by
observations using a telescope like ASKAP

The paper is organised as follows. We begin by summarising the
formalism for the HOD model, and introduce HOD modeling of 21cm
intensity fluctuations in \S~\ref{SB}. We discuss the potential
sensitivity of ASKAP to these fluctuations in \S~\ref{noise}. We then
present our forecast constraints on HOD parameters in
\S~\ref{constraints} and describe estimates of the HI mass function in
\S~\ref{HIMF}. We summarise our findings in \S~\ref{summary}. In our
numerical examples, we adopt the standard set of cosmological
parameters ~(Komatsu et al.~2009), with values of $\Omega_{\rm
  m}=0.24$, $\Omega_{\rm b}=0.04$ and $\Omega_Q=0.76$ for the matter,
baryon, and dark energy fractional density respectively, $h=0.73$, for
the dimensionless Hubble constant, and $\sigma_8=0.81$ for the
variance of the linear density field within regions of radius
$8h^{-1}$Mpc.

\section{Intensity mapping and the HI HOD}
\label{SB}

We begin by reviewing the
halo occupation distribution formalism for galaxies (e.g. Peacock \& Smith 2000; Seljak 2000; Scoccimarro
et al.~2001; Berlind \& Weinberg~2002; Zheng~2004) which we describe only briefly, referring the reader to the above papers for details.
The technique of surface brightness mapping will not allow resolution of individual galaxies, but rather the measurement of fluctuations in the surface brightness of unresulved galaxies. However we utilise a halo model formalism where galaxies are traced, rather than a form where the density field is a continuous function. This is because the HI is found in discrete galaxies, and treating individual galaxies allows us to explicitely calculate the HI mass weighted galaxy bias. The HOD model is constructed around the following simple
assumptions. First, one assumes that there is either zero or one
central galaxy that resides at the centre of each halo. Satellite
galaxies are then assumed to follow the dark matter distribution within
the halos. The mean number of satellites is typically assumed to
follow a power-law function of halo mass, while the number of
satellites within individual halos follows a Poisson (or some other)
probability distribution. The two-point correlation function on a scale $r$ can be
decomposed into one-halo ($\xi_{\rm 1h}$) and two-halo ($\xi_{\rm 2h}$) terms
\begin{equation}
\label{xi1}
\xi(r) = \left[1+\xi_{\rm 1h}(r)\right]+\xi_{2h}(r),
\end{equation}
corresponding to contributions to the correlation function from
galaxy pairs which reside in the same halo and in two different halos
respectively (Zheng~2004).

The 2-halo term can be computed as the halo correlation function
weighted by the distribution and occupation number of galaxies within
each halo. The 2-halo term of the galaxy power spectrum (PS) is
\begin{equation}
\label{2h}
P_{\rm gg}^{\rm 2h}(k) = P_{\rm m}(k)\left[\frac{1}{\bar{n}_{\rm g}}\int_{0}^{M_{\rm max}} dM \frac{dn}{dM} \langle N\rangle_{M} b(M) y_{\rm g}(k,M)\right]^2,
\end{equation}
where $P_{\rm m}$ is the mass PS and $y_{\rm g}$ is the
normalised Fourier transform of the galaxy distribution, which is assumed to follow a Navarro, Frenk \& White~(1997; NFW)
profile (see e.g. Seljak~2000; Zheng~2004). Here $\bar{n}_{\rm g}$ is the mean number density of galaxies. We
assume the Sheth Tormen~(1999) mass function $dn/dM$ using parameters
from Jenkins et al.~(2001) throughout this paper. To compute the halo bias $b(M)$ we use the Sheth, Mo and
Tormen~(2001) fitting formula. The quantity $M_{\rm max}$ is taken to
be the mass of a halo with separation $2r$. The 2-halo term for the
correlation function follows from
\begin{equation}
\label{x2h}
\xi_{2h}(r)=\frac{1}{2\pi^2}\int_0^\infty P_{\rm gg}^{\rm 2h}(k) k^2 \frac{\sin{kr}}{kr}dk.
\end{equation}

In real space the 1-halo term can be computed using (Berlind \& Weinberg~2002)
\begin{eqnarray}
\label{1h}
\nonumber
1+\xi_{1h}(r)&=&\frac{1}{2\pi r^2\bar{n}_{\rm g}^2}\\
&&\hspace{-20mm}\times\int_0^\infty dM\frac{dn}{dM}\frac{\langle N(N-1)\rangle_{M}}{2}\frac{1}{2R_{\rm vir}(M)}F^\prime\left(\frac{r}{2R_{\rm vir}}\right),
\end{eqnarray}  
where $\langle N(N-1)\rangle_{M}$ is the average number of galaxy pairs within halos of mass $M$. The distribution of multiple galaxies within a single halo is
described by the function $F^\prime(x)$, which is the differential probability that galaxy pairs are separated by a dimensionless distance $x\equiv r/R_{\rm vir}$. As is common in the
literature, we assume that there is always a galaxy located at the
center of the halo, and others are regarded as satellite galaxies.
The contribution to $F^\prime$ is therefore divided into pairs of galaxies
that do, and do not involve a central galaxy, and is computed assuming
that satellite galaxies follow the number-density distribution of an
NFW profile. With this assumption, the term in the integrand of equation~(\ref{1h}) reads
\begin{eqnarray}
\label{NN}
\nonumber
&&\hspace{-8mm}\langle N(N-1)\rangle_{M}F^\prime(x)\\
&&\hspace{-0mm} = \langle N-1\rangle_{M}F^\prime_{\rm cs}(x) + \frac{\langle(N-1)(N-2)\rangle_{M}}{2}F^\prime_{\rm ss}(x).
\end{eqnarray}
where $F^\prime(x)$ is the pair-number-weighted 
average of the central-satellite pair distribution $F^\prime_{\rm cs}(x)$
and the satellite-satellite pair distribution $F^\prime_{\rm ss}(x)$ (see,
e.g., Berlind \& Weinberg 2002; Yang et al. 2003; Zheng~2004),

\subsection{21cm intensity mapping of HI clustering}

The HOD method of estimating and modeling the clustering of HI
galaxies will not work at redshifts beyond $z\sim0.7$, where even the
most luminous galaxies will not be detectable in HI for the foreseeable
future. Instead, observations of surface brightness
fluctuations in 21cm intensity arising from the combined signal of a
large number of unresolved galaxies could be used to measure the
clustering of HI galaxies. Studies of 21cm surface brightness fluctuations
over a large volume will be made possible by the widefield
interferometers now coming on line, and will allow the HI properties
of galaxies to be studied over a greater range of redshifts. Indeed, it has been argued that lack
of identification of individual galaxies is an advantage when
attempting to measure the clustering of the HI emission, since by not
imposing a minimum threshold for detection, such a survey collects all
the available signal. This point is discussed in Pen et al.~(2008),
where the technique is also demonstrated via measurement of the
cross-correlation of galaxies with unresolved 21cm emission in the
local Universe.

The situation is analogous to mapping of the three-dimensional
distribution of cosmic hydrogen during the reionization era through
the 21cm line~(Furlanetto, Oh \& Briggs~2007; Barkana \& Loeb~2007). Several
facilities are currently being constructed to perform this experiment
(including MWA~\footnote{http://www.haystack.mit.edu/ast/arrays/mwa/},
LOFAR~\footnote{http://www.lofar.org/}, PAPER
~\footnote{http://astro.berkeley.edu/$\sim$dbacker/EoR/},
21CMA~\footnote{http://web.phys.cmu.edu/$\sim$past/}) and more ambitious
designs are being planned
(SKA~\footnote{http://www.skatelescope.org/}). During the epoch of
reionization, the PS of 21cm brightness fluctuations is shaped mainly
by the topology of ionized regions. However the situation is expected
to be simpler following reionization of the intergalactic medium (IGM;
$z\la6$) -- when only dense pockets of self-shielded hydrogen, such as
damped Ly$\alpha$ absorbers (DLA) and Lyman-limit systems (LLS)
survive ~(Wyithe \& Loeb~2008; Chang et al.~2007; Pritchard \&
Loeb~2008). These DLA systems are thought to be the high redshift
equivalents of HI rich galaxies in the local Universe (Zwaan et al.~2005b).  We do not expect 21cm self absorption to impact the level of
21cm emission. This conclusion is based on 21cm absorption studies
towards damped Ly$\alpha$ systems at a range of redshifts between
$z\sim0$ and $z\sim3.4$, which show optical depths to absorption of
the back-ground quasar flux with values less than a few percent
(Kanekar \& Chengalur 2003; Curran et al. 2007).  Moreover, damped
Ly$\alpha$ systems have a spin temperature that is large relative to
the temperature of the cosmic microwave background radiation, and will
therefore have a level of emission that is independent of the kinetic
gas temperature (e.g. Kanekar \& Chengalur 2003). Thus the intensity of 21cm emission can be directly related to the column density of HI.

\subsection{Modeling the power spectrum of 21cm fluctuations }

As mentioned above, low spatial resolution observations could be used
to detect surface brightness fluctuations in 21cm emission from the
cumulative sum of HI galaxies, rather than from individual sources of
emission. Here the PS is a more natural observable than
the correlation function, since a radio interferometer records visibilities
that directly sample the PS. In the linear regime the 21cm PS follows directly from the PS of fluctuations in mass
$P_{\rm m}(k)$ (Wyithe \& Loeb~2009)
\begin{equation}
\label{DLAPS1}
P_{\rm HI}(k)\approx\mathcal{T}_{\rm b}^2x_{\rm HI}^2\langle b\rangle_{M}^2P_{\rm m}(k),
\end{equation}
where $\mathcal{T}_{\rm b}=23.8\left[{(1+z)}/{10}\right]^\frac{1}{2}\,$mK is the brightness temperature contrast between the mean IGM and the CMB at redshift $z$, and $\langle b\rangle_{M}$ is the HI mass weighted halo bias. Note that we have used the subscript $_{\rm HI}$ rather than the more usual $_{21}$ in order to reduce confusion with the subscripts for the 1-halo and 2-halo PS terms. The fraction of hydrogen that is neutral is 
described by the parameter $x_{\rm HI}\equiv\Omega_{\rm
 HI}/(0.76\Omega_{\rm b})$. We assume $x_{\rm HI}=0.01$ (corresponding to $\Omega_{\rm HI}\sim3\times10^{-4}$, Zwaan et al.~2005a) throughout this paper.
The resulting PS is plotted in the right panel of Figure~\ref{fig1} (dashed line).  The constant
$\mathcal{T}_{\rm b}$ hides the implicit assumptions that the 21cm
emission from the galaxies is not self absorbed, and that the spin
temperature is much larger than the temperature of the cosmic
microwave background. On small scales a model is needed to relate HI mass to halo mass. To achieve this we modify the HOD formalism as outlined below.

\subsubsection{HOD model for 21cm  fluctuations}

Since surface brightness fluctuations depend on the total HI mass within a halo rather than on number counts of individual galaxies, the number of galaxies, and the number of galaxy pairs per halo in the HOD formalism need to be weighted by the HI mass per galaxy. In analogy with the HOD formalism, we distribute this mass between central and satellite galaxies. We define $\langle M_{\rm HI,c}\rangle_{M}$ and $\langle M_{\rm HI,s}\rangle_{M}$ to be the mean HI mass of central galaxies and of the combined satellite galaxies within a halo of mass $M$ respectively.

To compute the 2-halo PS, we replace $\langle N\rangle_{M}$ in equation~(\ref{2h}) with the mean value of the total HI mass in a halo of mass $M$, i.e. $\langle M_{\rm HI} \rangle_{M}= \langle M_{\rm HI,c} \rangle_{M} + \langle M_{\rm HI,s} \rangle_{M}$, yielding  
\begin{eqnarray}
\label{Pgg}
\nonumber
P_{\rm HI,gg}^{\rm 2h}(k) = \mathcal{T}_{\rm b}^2x_{\rm HI}^2P_{\rm m}(k)\\
&&\hspace{-33mm}\times\left[\frac{1}{\bar{\rho}_{\rm HI}}\int_{0}^{M_{\rm max}} dM \frac{dn}{dM} \langle M_{\rm HI}\rangle_{M} b(M) y_{\rm g}(k,M)\right]^2,
\end{eqnarray}
where, $\bar{\rho}_{\rm HI}$ is the mean density of HI contributed by all galaxies in the IGM. The 2-halo term $\xi_{2h,\rm HI}(r)$ follows from substitution into equation~(\ref{x2h}).

To compute the 1-halo term we again weight the number of galaxies by their HI mass. In difference from the calculation of the 1-halo term for galaxy clustering, the distribution of satellite masses will be important in addition to the number. This aspect of the HOD modeling will require simulation for a proper treatment (e.g. Bagla \& Khandai~2009). However for the purposes of our analysis it is sufficient to assume that most of the satellite HI for a halo mass $M$ is contained within satellites of similar mass (as would be the case for a steep power-law mass function with a lower cutoff for example). We therefore further define $\langle m_{\rm HI,s}\rangle_{M}$ to be the mean HI mass of satellite galaxies within a halo of mass $M$. The coefficients in equation~(\ref{NN}) are then modified to yield
\begin{equation}
\langle (N-1)\langle m_{\rm HI,s}\rangle_{M}\times\langle M_{\rm HI,c}\rangle_{M}\rangle= \langle M_{\rm HI,c}\rangle_{M}\langle M_{\rm HI,s}\rangle_{M}
\end{equation}
and
\begin{equation}
\frac{\langle(N-1)\langle m_{\rm HI,s}\rangle_{M}\times (N-2)\langle m_{\rm HI,s}\rangle_{M}\rangle}{2}=\frac{\langle M_{\rm HI,s}\rangle_{M}^2}{2}
\end{equation}
respectively. To calculate this second term we have noted that for a Poisson distribution of galaxies $\langle(N-1)(N-2)\rangle_{M} = \langle(N-1)\rangle^2_{M}$, and have assumed that $\langle M_{\rm s}\rangle_{M} = \langle (N-1)\rangle_{M}\times \langle m_{\rm s}\rangle_{M}$. The modified expression for the 1-halo term therefore becomes\footnote{Note that the correlation function has dimensions of mK, and therefore does not have the usual interpretation of probability above random for finding a galaxy pair of separation $r$.}
\begin{eqnarray}
\label{1hM}
\nonumber
&&\hspace{-7mm}\mathcal{T}_{\rm b}^2x_{\rm HI}^2+\xi_{1h,\rm HI}(r)=\\
\nonumber
&&\hspace{-2mm}\frac{\mathcal{T}_{\rm b}^2x_{\rm HI}^2}{2\pi r^2\bar{\rho}_{\rm HI}^2}\int_0^\infty \frac{dn}{dM} \left[ \langle M_{\rm HI,c}\rangle_{M}\langle M_{\rm HI,s}\rangle_{M} F^\prime_{\rm cs}\left(\frac{r}{2R_{\rm vir}}\right)\right.\\
 &&\hspace{15mm}+ \left.\frac{\langle M_{\rm HI,s}\rangle_{M}^2}{2}F^\prime_{\rm ss}\left(\frac{r}{2R_{\rm vir}}\right)   \right]   \frac{1}{R_{\rm vir}(M)}dM.
\end{eqnarray}  
The correlation function follows from $\xi_{\rm \rm HI}(r)=[\mathcal{T}_{\rm b}^2x_{\rm HI}^2+\xi_{\rm 1h,\rm HI}(r)]+\xi_{\rm 2h,\rm HI}(r)$.

In order to evaluate this expression the HI mass occupation of a halo of mass $M$ must be parameterised, and is obviously quite uncertain. For illustration, we choose the following polynomial form, with a minimum halo mass ($M_{\rm min}$) and characteristic scale ($M_1$) where satellites contribute HI mass that is comparable to the central galaxy,
\begin{eqnarray}
\label{Mparam}
\nonumber
\langle M_{\rm HI,c}\rangle_{M} &\propto& M^{\gamma_{\rm c}} \hspace{20.5mm}\mbox{if}\hspace{5mm}M>M_{\rm min}\\
&=&0\hspace{25mm}\mbox{otherwise}.
\end{eqnarray}
and 
\begin{eqnarray}
\nonumber
\langle M_{\rm HI,s}\rangle_{M} &=& M_{\rm HI,c}\left(\frac{M}{M_1}\right)^{\gamma_{\rm s}}   \hspace{4.5mm}\mbox{if}\hspace{5mm}M>M_{\rm min}\\
&=&0\hspace{25mm}\mbox{otherwise}.
\end{eqnarray}
The average HI mass within a halo of mass $M>M_{\rm min}$ is therefore 
\begin{equation}
\label{total}
\langle M_{\rm HI}\rangle_{M} = \langle M_{\rm HI,c}\rangle_{M}+\langle M_{\rm HI,s}\rangle_{M}\propto M^{\gamma_{\rm c}}\left[1+\left(\frac{M}{M_1}\right)^{\gamma_{\rm s}}\right]
\end{equation}
Note that the constant of proportionality in equations~(\ref{Mparam})
and (\ref{total}) is not specified but cancels with the same factor in
$\rho_{\rm HI}$ in equations~(\ref{Pgg}) and (\ref{1hM}). From
experience of the galaxy HOD there will be degeneracy between the
parameters $M_1$ and $\gamma_{\rm s}$. We therefore make the
simplification of setting $\gamma_{\rm c}=\gamma_{\rm s}\equiv\gamma$
in our parameterisation for the remainder of this paper. The left
panel of Figure~\ref{fig1} shows the real-space correlation function
at $z=0.7$ for an HOD model with parameters $\gamma=0.5$, $M_{\rm
  min}=10^{11}$M$_\odot$ and $M_1=10^{13}$M$_\odot$. This model serves as our fiducial case throughout this paper, and is
motivated by the parameters derived from estimates for HIPASS galaxies  (Wyithe et al.~2009). In particular we note the value of $\gamma=0.5$ which is smaller than unity. This value encapsulates the assumption that smaller halos have more HI, and agrees with the conventional wisdom that galaxy clusters are HI poor. In the local Universe $\gamma\sim0.5$ is found to describe the relation between HI and dynamical masses of HIPASS galaxies (Wyithe et al.~2009). Aside from this motivation the fiducial model is otherwise arbitrary.

\subsubsection{Redshift space modeling of the 21cm power spectrum}

\begin{figure*}
\centerline{\epsfxsize=4.9in \epsfbox{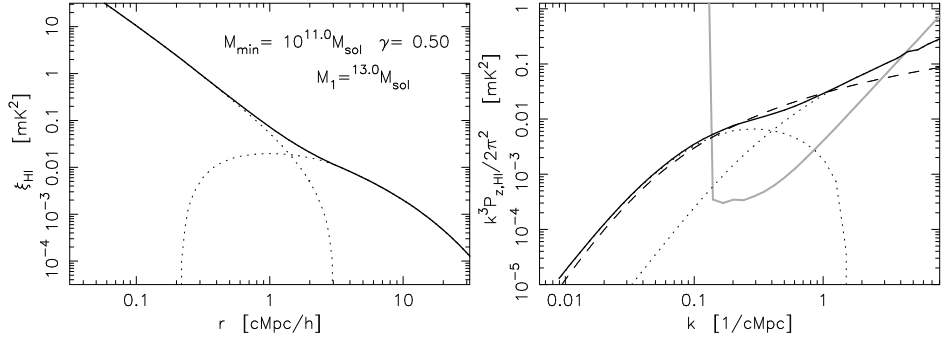}} 
\caption{
An example of the correlation function and PS of intensity fluctuations for a hypothetical HOD model at $z=0.7$. {\em Left-hand panel:} the model correlation function. {\em Right-hand panel:} The corresponding spherically averaged redshift space PS (solid line). In each case the 1-halo and 2-halo terms are plotted as dotted curves. For comparison we plot the spherically averaged sensitivity  within pixels of width $\Delta k=k/10$ for an a radio interferometer resembling the design of ASKAP (thick gray line). We also show the real-space 21cm PS assuming a linear mass-density PS (dashed line). 
For calculation of observational noise an integration of 3000 hours was assumed, with a multiple primary beam total field of view corresponding to $30(1+z)^2$ square degrees (see text for details). The cutoff at large scales is due to foreground removal within a finite frequency band-pass.}
\label{fig1}
\end{figure*}

Since a radio interferometer directly measures the 3 dimensional
distribution of 21cm intensity it is more powerful to work in redshift
space, where line-of-sight infall (Kaiser~1987) can be used to break
the degeneracy between neutral fraction and galaxy bias
(Wyithe~2008). In addition to gravitational infall the shape of the
redshift space PS will be complicated by peculiar motions of galaxies
within groups or clusters, which produce the so-called fingers-of-god
in the redshift space correlation function. In the case of 21cm
fluctuations, the internal velocities of HI in galaxies will also
contribute to the fingers-of-god. In this paper we use the combination
of the real-space HOD 21cm PS
\begin{equation}
P_{\rm R,\rm HI}(k) = 4\pi\int dr \xi_{\rm \rm HI}(r)\frac{\sin{kr}}{kr} r^2 dr,
\end{equation}
and the dispersion model to estimate the redshift space PS including
these effects. The dispersion model is written
\begin{equation}
P_{\rm z,\rm HI}(k_{\rm perp},k_{\rm
  los}) = P_{\rm R,HI}(k)(1+\beta\mu^2)^2(1+k^2\sigma_k^2\mu^2/2)^{-1}
\end{equation}
where $\mu$ is the cosine of the angle between the line-of-sight and
the unit-vector corresponding to the direction of a particular mode,
$k_{\rm perp}=k\mu$, $k_{\rm los}=k\sqrt{1-\mu^2}$, $\beta=\Omega_{\rm
  m}^{0.6}/\langle b\rangle_{\rm M}$ and $\langle b\rangle_{\rm M}$ is
the average HI mass weighted halo bias. The quantity $\sigma_k$ is a
constant which describes a ``typical'' velocity dispersion for
galaxies and parametrises the prominence of the fingers-of-god.
Simulations indicate a value of $\sigma_k\sim650/(1+z)$km$\,$s$^{-1}$
(Lahav \& Suto~2004). We note that the redshift space PS could have
been generated through a 2-d Fourier transform of the redshift space
correlation function computed within the HOD model using the formalism
described in Tinker (2007), allowing additional constraints on HOD
parameters to be placed based on the prominence of the fingers-of-god
as in Wyithe et al.~(2009). In particular the assumption of $\gamma<1$
as well as the potential lack of HI in clusters would lead to reduced
prominance of the fingers of god. However in the absence of current
data we have taken the simpler approach of parameterising the
fingers-of-god using $\sigma_k$.

\begin{figure*}
\centerline{\epsfxsize=4.9in \epsfbox{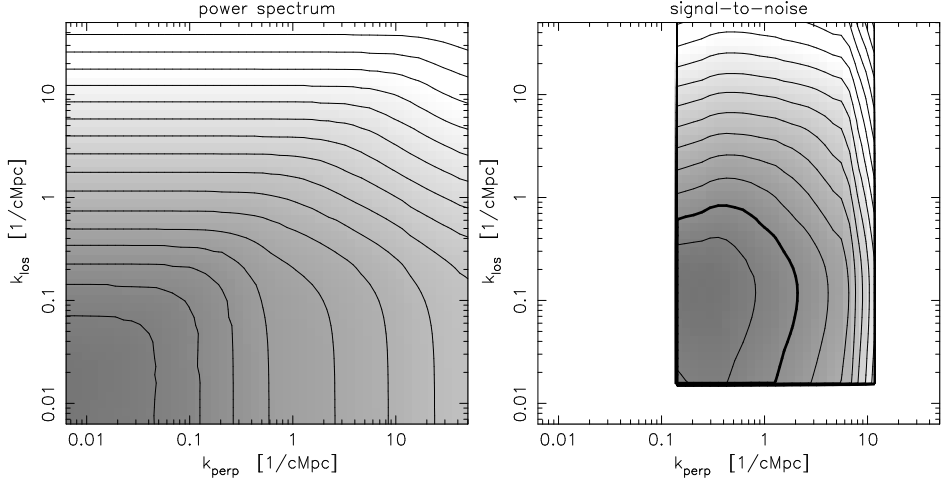}} 
\caption{
The redshift space PS of intensity fluctuations corresponding to the example in Figure~\ref{fig1}. {\em Left-hand panel:} The redshift-space PS assuming a dispersion model with $\sigma_k=650/(1+z)$km$\,$s$^{-1}$. {\em Right-hand panel:} Contours of the signal-to-noise (separated by factors of $\sqrt{10}$) within pixels of width $\Delta k=k/10$. The thick contour corresponds to a signal-to-noise per pixel of unity.  For calculation of observational noise an integration of 3000 hours was assumed, with a multiple primary beam total field of view corresponding to $30(1+z)^2$ square degrees (see text for details). The cutoffs at large and small scales perpendicular to the line-of-sight are due to the lack of short and long baselines respectively. The cutoff at large scales along the line-of-sight is due to foreground removal within a finite band-pass.}
\label{fig2}
\end{figure*}

\begin{figure*}
\centerline{\epsfxsize=4.9in \epsfbox{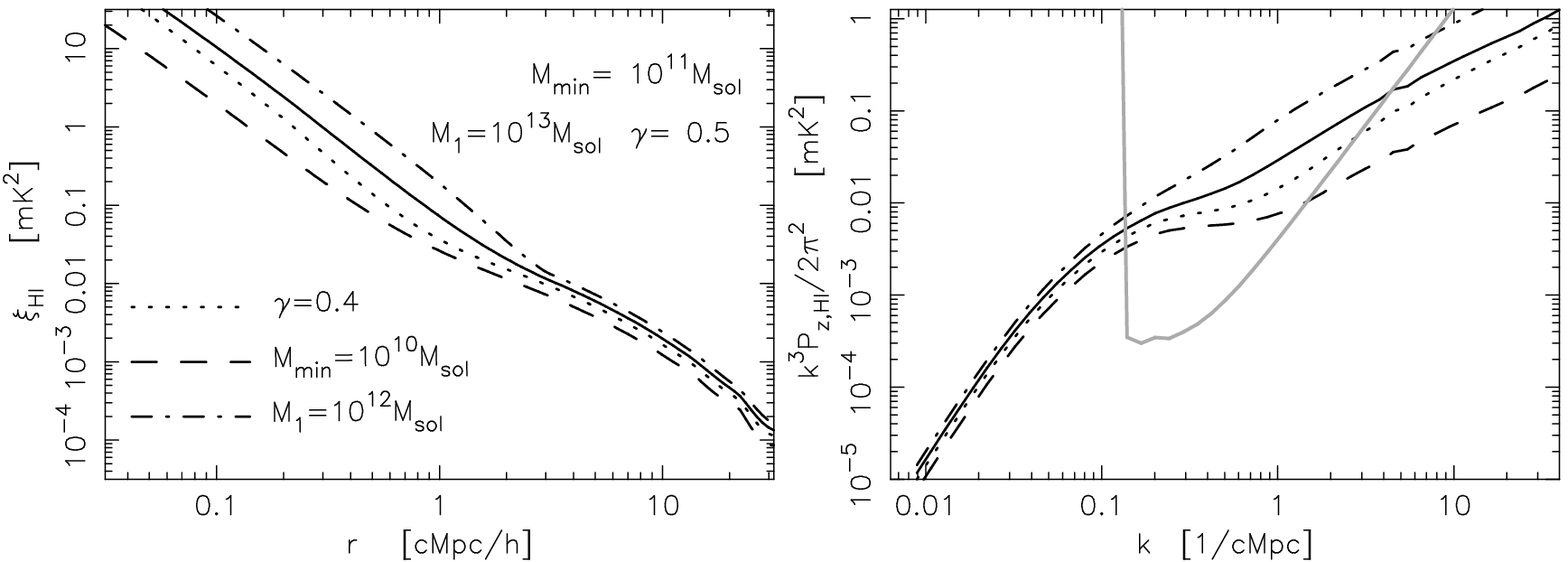}} 
\caption{
Examples of correlation functions ({\em left panel}) and spherically averaged redshift space PS ({\em right panel}) of intensity fluctuations for different HOD models at $z=0.7$. The solid lines repeat
the fiducial model from Figure~\ref{fig1} (parameters listed in the top-right corner of the left panel). For comparison, the dotted, dashed and
dot-dashed lines show variations on this model, with $\gamma=0.4$,
$M_{\rm min}=10^{10}$M$_\odot$ and $M_{1}=10^{12}$M$_\odot$
respectively (with the remaining parameters set to their fiducial
values in each case). 
For comparison with the PS, we also plot the spherically averaged sensitivity  within pixels of width $\Delta k=k/10$ for an a radio interferometer resembling the design of ASKAP (right panel, thick gray line). }
\label{fig3}
\end{figure*}

The left panel of Figure~\ref{fig2} shows the resulting redshift space
PS for the fiducial model. The large scale motions induced by infall into overdense regions
can be seen as an extension of the PS along the line-of-sight at small
$k$, while the fingers of god are manifest as a compression at large
$k$. The right panel of Figure~\ref{fig1} shows the corresponding spherically
averaged redshift space PS (solid line)
\begin{equation}
P^{\rm sph}_{\rm z,\rm HI}(k) = \int P_{\rm z,\rm HI}(k_{\rm perp},k_{\rm los}) dk_{\rm perp}dk_{\rm los}.
\end{equation}
For comparison, the dotted lines in the right hand panel of Figure 2
show the 1-halo and 2-halo contributions to the spherically averaged
redshift space PS.
 
The spherically averaged redshift space PS can be compared with the
linear real-space 21cm PS estimated based on the neutral fraction
$x_{\rm HI}=0.01$ and the linear mass PS (dashed line,
equation~\ref{DLAPS1}).  On large (linear) scales the spherically
averaged PS is larger in redshift space than in real space. This is
analogous to the excess power seen in redshift space clustering of
galaxy surveys (Kaiser~1987), and is due to an increase in the 21cm
optical depth owing to velocity compression (Barkana \& Loeb~2005)
towards high density regions. On small scales there is excess power
above the linear theory expectation owing to the inclusion of the
non-linear 1-halo term. The 21cm PS shows a steepening at large $k$
owing to the mass weighting in the 1-halo term. This steepening is
also seen in the simulations of Bagla \& Khandai~(2009).

\subsubsection{variation of the 21cm PS with HOD parameters}
Figure~\ref{fig3} illustrates the sensitivity of the clustering and
the 21cm PS to variations in the HOD parameters. The solid lines repeat
the fiducial model from Figure~\ref{fig1}. For comparison, the dotted, dashed and
dot-dashed lines show variations on this model, with $\gamma=0.4$,
$M_{\rm min}=10^{10}$M$_\odot$ and $M_{1}=10^{12}$M$_\odot$
respectively (with the remaining parameters set to their fiducial
values in each case). On the largest scales the clustering amplitude
is most sensitive to $M_{\rm min}$ (dashed lines), which enables the
typical host halo mass of HI to be measured from the PS
amplitude (Wyithe~2008). Lowering the value of $M_{\rm min}$ (while keeping $M_1$
fixed) implies a smaller fraction of HI in satellites, and hence a
relative decrease of power on small scales. A smaller value of
$\gamma$ also leads to a relative decrease of power on small scales
because the flatter power-law preferentially places mass in the more
common low mass halos (with $M<M_1$), and so lowers the fraction of HI
in satellites (dotted lines). Conversely, a smaller value of $M_1$ leads
to a larger fraction of HI in satellite systems, and hence an increase
of small scale power (dot-dashed lines). 

The variation in shape and amplitude of the 21cm PS implies that parameter values for a particular HOD model could be constrained if the PS were measured with sufficient signal-to-noise. In the remainder of this
paper we therefore first discuss the sensitivity of a radio interferometer to the 21cm PS, and then estimate the corresponding constraints on the 5 parameters in our HOD
model that could be placed using observations of 21cm intensity
fluctuations.

\section{Sensitivity to the 21cm PS}
\label{noise}

In this paper we estimate the ability of
a telescope like ASKAP to measure the
clustering of 21cm intensity fluctuations, and hence to estimate HOD parameters, and the total HI content of the Universe. The latter
quantity, which is not available from the clustering of resolved
galaxies, could be used to bridge the gap in measurements of
$\Omega_{\rm HI}$ (the cosmic density of HI relative to the
critical density) between the local Universe where this quantity can
be determined from integration of the HI mass function, and $z\ga2$
where it is measured from the column density through counting of
damped Ly$\alpha$ absorbers. 

To compute the sensitivity $\Delta P_{\rm HI}(k_{\rm perp},k_{\rm
  los})$ of a radio-interferometer to the 21cm PS, we follow the
procedure outlined by ~McQuinn et al.~(2006) and Bowman, Morales \&
Hewitt~(2007) [see also ~Wyithe, Loeb \& Geil~(2008)]. The important
issues are discussed below, but the reader is referred to these papers
for further details. The uncertainty comprises components due to
the thermal noise, and due to sample variance within the finite volume
of the observations.  We also include a Poisson component due to the
finite sampling of each mode~(Wyithe~2008), since the
post-reionization 21cm PS is generated by discrete clumps rather than
a diffuse IGM.  We consider
a telescope based on ASKAP. This
telescope is assumed to have 36 dish antenna with a density distributed as
$\rho(r)\propto r^{-2}$ within a diameter of 2km. The antennae are
each 12m in diameter, and being dishes are assumed to have physical and effective collecting areas that are equal. We assume that foregrounds can be removed over
$80$MHz bins, within a bandpass of $300$MHz [based on removal within
1/4 of the available bandpass (McQuinn et al.~2006)]. Foreground removal
therefore imposes a minimum on the wave-number accessible of
$k\sim0.02[(1+z)/1.5]^{-1}$Mpc$^{-1}$, although access to the large
scale modes is actually limited by the number of short baselines available. An
important ingredient is the angular dependence of the number of modes
accessible to the array~(McQuinn et al.~2006). ASKAP is designed to
have multiple primary beams facilitated by a focal plane phased
array. We assume 30 fields are observed simultaneously for 3000 hr
each, yielding $\sim30(1+z)^2$ square degrees [where the factor of $(1+z)^2$ originates from the frequency dependence of the primary beam]. The signal-to-noise for observation of the PS in the left panel of Figure~\ref{fig2} is shown in the right panel of
Figure~\ref{fig2}. A telescope like ASKAP would be most sensitive to
modes of $k_{\rm perp}\sim0.1-1$Mpc$^{-1}$ and $k_{\rm los}\sim0.03-0.3$Mpc$^{-1}$.
The spherically averaged signal-to-noise (within bins of $\Delta k=k/10$) is shown in the right panels of Figures~\ref{fig1} and \ref{fig3} (grey curves). Comparison of the noise curve with the variability of the PS amplitude and shape among different HOD models for the 21cm PS (Figure~\ref{fig3}) indicates that a telescope like ASKAP would be sufficiently sensitive to generate constraints on the HOD. Moreover, the spatial scale on which the array would be most sensitive corresponds to wave-numbers where we expect 1-halo and 2-halo contributions to be comparable, indicating that such observations may constrain HOD model parameters. This statement is quantified in the next section.

\section{Constraints on HOD parameters from 21cm intensity mapping}
\label{constraints}

\begin{figure*}
\centerline{\epsfxsize=6.9in \epsfbox{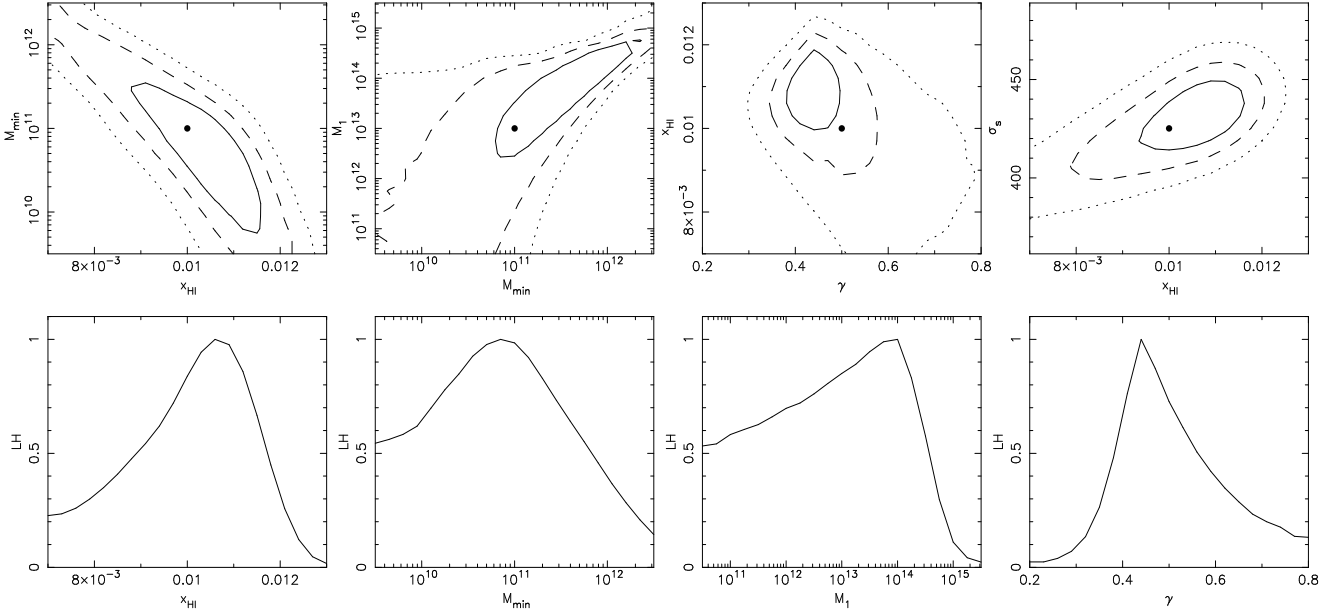}} 
\caption{
Example of forecast constraints on HOD model parameters from 21cm intensity fluctuations, assuming a 3000hr integration of a 30$(1+z)^2$ square degree field with an array based on the ASKAP design centered on $z=0.7$ (further details in the text). The \textit{upper panels} show contours of the likelihood in 2-d projections of the 5-parameter space used for the HOD modeling of 21cm intensity fluctuations, while the \textit{lower panels} show the marginalised likelihoods on individual parameters. Here prior probabilities on $\gamma$, $\log{x_{\rm HI}}$, $\log{M_{\rm min}}$ and $\log{M_{\rm 1}}$ are assumed to be constant. The contours are placed at 60\%, 30\% and 10\% of the peak likelihood. The position of the dot indicates the peak likelihood in the 5-dimensional parameter space (i.e. the input model).}
\label{fig4}
\end{figure*}

Based on our estimate of the sensitivity to the 21cm PS we forecast the ability of ASKAP to constrain the HI HOD. 
To begin, we assume the fiducial model $P_{\rm z,\rm HI}^{\rm true}(k_{\rm
  perp},k_{\rm los})$, as shown in Figure~\ref{fig2}, and estimate the
accuracy with which the parameters could be inferred. Our HOD model
for the 21cm PS has five parameters, $M_{\rm min}$,
$M_{1}$, $\gamma$, $x_{\rm HI}$ and $\sigma_k$. For combinations of
these parameters ($M_{\rm min}$,
$M_{1}$, $\gamma$, $x_{\rm HI}$, $\sigma_k$) that differ from the fiducial case ($10^{11}$M$_\odot$, $10^{13}$M$_\odot$, 0.5, 0.01, $650/(1+z)$km/s), we compute a trial model for the real space correlation function. We then use this to
calculate the chi-squared of the difference between the fiducial and the trial models
\begin{eqnarray} 
\chi^2=&&\\
\nonumber
&&\hspace{-15mm}\sum_{k_{\rm perp}}\sum_{k_{\rm los}}\left(\frac{P^{\rm true}_{\rm z,\rm HI}-P_{\rm z,\rm HI}(k_{\rm perp},k_{\rm los}|M_{\rm min},M_{1},\gamma,x_{\rm HI},\sigma_k,\sigma_8)}{\Delta P_{\rm z,\rm HI}(k_{\rm perp},k_{\rm los})}\right)^2,
\end{eqnarray} 
and hence find the likelihood
\begin{equation}
\label{LHHOD}
\mathcal{L}(M_{\rm min},M_{1},\gamma,x_{\rm HI},\sigma_k) = \int d\sigma^\prime_8 \frac{dp}{d\sigma^\prime_8} \exp{\left(-\frac{\chi^2}{2}\right)}.
\end{equation}
The uncertainty introduced through imperfect knowledge of the PS amplitude (which is proportional
to the normalization of the primordial PS, $\sigma_8$) is degenerate
with $x_{\rm H}$ (Wyithe~2008). For this reason the uncertainty in $\sigma_8$ has been explicitly included in equation~(\ref{LHHOD}). We assume a Gaussian distribution $dp/d\sigma_8$ for $\sigma_8$ with $\sigma_8=0.81\pm0.03$ (Komatsu et al.~2009).

Figure~\ref{fig4} shows an example of forecast constraints on HOD model parameters for a telescope like ASKAP, assuming a 3000hr integration of a single pointing $[\sim30(1+z)^2$ square degrees] centered on $z=0.7$. Results are presented in the upper panels of Figures~\ref{fig4} which shows contours of the likelihood in 2-d projections of this 5-parameter space. Here
prior probabilities on $\log{x_{\rm HI}}$, $\log{M_{\rm min}}$,
$\log{M_{\rm 1}}$, $\gamma$ and $\sigma_k$ are assumed to be
constant. The contours are placed at 60\%, 30\% and 10\% of the peak
likelihood. The lower
panels show the marginalised likelihoods on the individual parameters
$x_{\rm HI}$, $M_{\rm min}$, $M_{\rm 1}$ and $\gamma$. 

A deep integration of a single pointing for a telescope like ASKAP
would place some constraints on the minimum mass (the projected
uncertainty on $M_{\rm min}$ is $\sim0.5$ dex), and measure the
relationship between HI and halo mass (a $\sim 20\%$ constraint on
$\gamma$). In addition to these constraints on the halo occupation
distribution of HI, observations of the 21cm PS would also provide a
measurement of the global neutral fraction (or equivalently
$\Omega_{\rm HI}$), which would be constrained with a relative
uncertainty of 20\% at $z\sim0.7$. This indicates that 21cm intensity
fluctuations could be used to measure the evolution of $\Omega_{\rm
  HI}$ from $z\sim1$ to the present day, even though ASKAP will not
detect individual galaxies at $z\ga0.7$ (Johnston et
al.~2007). Measurement of $\Omega_{\rm HI}$ based on the redshift space
PS would be complimentary to the detection of individual rare peaks,
which could facilitating a direct estimate of the cosmic HI mass
density (Bagla \& Khandai~2009).

\section{HI content and the HI mass function}
\label{HIMF}

\begin{figure*}
\centerline{\epsfxsize=5.9in \epsfbox{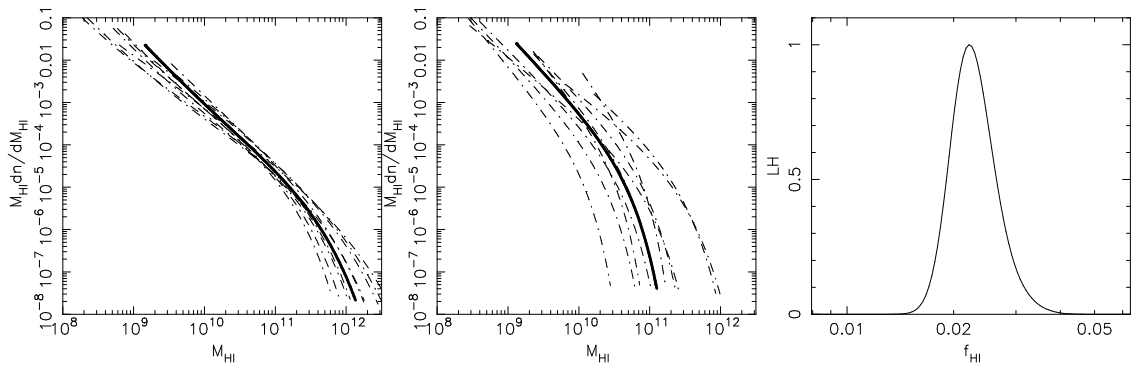}} 
\caption{
Examples of the range for the total halo (left) and central galaxy (central panel) HI mass functions. In addition to the fiducial case (thick lines, corresponding to the model in Figure~\ref{fig1}), ten HI mass functions are shown in each case with parameters drawn from the probability distribution for HOD parameters in Figure~\ref{fig4} (dot-dashed lines). In the right panel we show forecast for corresponding constraints on $f_{\rm HI}$. A 3000hr integration of a 30$(1+z)^2$ square degree field with an array based on the ASKAP design centered on $z=0.7$ (corresponding to Figure~\ref{fig4}) was assumed. The prior probability on $\log{f_{\rm HI}}$ was assumed to be constant. }
\label{fig5}
\end{figure*}

The combination of measurements for $x_{\rm HI}$ and the HOD parameters $\gamma$, $M_{\rm min}$ and $M_1$ indicates that the HI mass function (summing both central and satellite galaxies) could be approximated from the HOD using
\begin{equation}
\frac{dn}{dM_{HI}} = \frac{dn}{dM}\frac{dM}{dM_{\rm HI}}.
\end{equation}
where $M_{\rm HI}=C M^\gamma (1+(M/M_1)^\gamma)$ and the constant $C$ is evaluated from  
\begin{equation}
C = x_{\rm HI} \frac{0.76(\Omega_{\rm b}/\Omega_{\rm M})\rho_{\rm m}}{\int_{M_{\rm min}}^\infty M^\gamma \left(1+\left(\frac{M}{M_1}\right)^\gamma\right)\frac{dn}{dM}dM}.
\end{equation}
The left panel of Figure~\ref{fig5} shows the mass function for the
fiducial model in Figure~\ref{fig1} (thick grey curve) as well as ten
HI mass functions computed assuming parameters drawn at random from
the joint probability distribution $[\propto M_{\rm
  min}^{-1}M_{1}^{-1}x_{\rm HI}^{-1}\mathcal{L}(M_{\rm
  min},M_{1},\gamma,x_{\rm HI},\sigma_k)]$, projections of which are
shown in Figure~\ref{fig4}. While the values over which the HI mass
range extends in these realisations shows some variability, the
possibility of constraints on $\gamma$ and $x_{\rm HI}$ implied by
Figure~\ref{fig4} mean that the overall shape of the HI mass function
could be quite well constrained by observation of a redshifted 21cm
PS. In the central panel of Figure~\ref{fig5} we show the
corresponding the HI mass functions for central galaxies [obtained by
instead substituting $M_{\rm HI}=C M^\gamma$]. In
this case the range of realisations is much larger, which can be
traced to the degeneracy between $M_{\rm min}$ and $M_1$ seen in
Figure~\ref{fig4}.

In addition to the HI mass function, it would also be possible to constrain the fraction of hydrogen within galaxies that is in atomic form. This number is given by 
\begin{equation}
  f_{\rm HI}=\frac{x_{\rm HI}}{F_{\rm col}(M_{\rm min})},
\end{equation}
where $F_{\rm col}(M_{\rm min})$ is the fraction of dark-matter that
is collapsed in halos more massive than $M_{\rm min}$. From the upper
left panel of Figure~\ref{fig4} we see that there is a degeneracy
between $x_{\rm HI}$ and $M_{\rm min}$. Larger neutral fractions
correspond to lower values of $M_{\rm min}$ and hence larger collapsed
fractions. As a result the ratio $f_{\rm HI}$ would be very well
constrained, as shown by the likelihood distribution in the right hand panel of
Figure~\ref{fig5} (which is based on the distributions in
the left panel of Figure~\ref{fig5}). The evolution of $f_{\rm}$, which can also be measured
locally from clustering with a value of $f_{\rm HI}=10^{-1.4\pm0.4}$
(Wyithe, Brown, Zwaan \& Meyer~2009) will provide an important
ingredient for studies of the role of HI in star formation.

\section{Summary}
\label{summary}

Due to the faintness of HI emission from individual galaxies, even
deep HI surveys will be limited to samples at relatively low redshift 
($z\la0.7$) for the next decade. However these surveys will
be able to detect fluctuations in 21cm intensity produced by the
ensemble of galaxies out to higher redshifts, using observational
techniques that are analogous to those being discussed with respect to
the reionization epoch at $z\ga6$ (e.g. Furlanetto, Oh \&
Briggs~2006).  As a result, studies of HI galaxy clustering could be
extended to redshifts beyond those where individual HI galaxies can be
identified through the use of 21cm intensity fluctuations.  To investigate this possibility we have described an
approximate model for the power spectrum of 21cm fluctuations, which
is based on the halo occupation distribution formalism for galaxy
clustering. Our goal for this paper has been to use this model
to estimate the expected amplitude and features of the 21 cm
power-spectrum, rather than to present a detailed method for
extracting the halo occupation of HI from an observed
power-spectrum. This latter goal would require numerical simulations
(e.g. Bagla \& Khandai 2009).

To frame our discussion we have made forecasts for ASKAP, specifically
with respect to the use of the 21cm power spectrum as a probe of the
occupation of HI in dark matter halos. We have chosen
$z=0.7$ for our estimates, which is the redshift at which individual
galaxies are no longer detectable with ASKAP in deep integrations. We
have shown that a telescope based on the design of ASKAP will have
sufficient sensitivity to yield estimates of the HI halo
occupation. Because 21cm intensity fluctuations combine the
integrated HI from all galaxies (not just those detected as individual
sources), the clustering amplitude is proportional to the total HI
content of the Universe. We find that an array with the specifications
of ASKAP could yield estimates of the global HI density which have a
relative accuracy of $\sim20\%$. Clustering measurements in 21cm
surface brightness could therefore be used to make measurements of the
global HI content in the currently unexplored redshift range between
the local Universe, and surveys for damped Ly$\alpha$ absorbers in the
higher redshift Universe.

The cosmic star-formation rate has declined by more than an order of
magnitude in the past 8 billion years (Lilly et al. 1996, Madau et al.
1996). Optical studies paint a somewhat passive picture of
galaxy formation, with the stellar mass density of galaxies gradually
increasing and an increasing fraction of stellar mass mass ending up
within red galaxies that have negligible star-formation (e.g., Brown
et al. 2008). On the other hand, the combination of direct HI observations 
at low redshift (Zwaan et al. 2005; Lah et al 2007) and damped Ly$\alpha$
absorbers in the spectra of high-redshift QSOs (Prochaska et al. 2005)
show that the neutral gas density has remained remarkably constant
over the age of the universe.  The evolutionary
and environmental relationships between the neutral gas which provides
the fuel for star formation and the stars that form are central to
understanding these and related issues. The study of the halo
occupation distribution of HI based on 21cm fluctuations has the potential to allow these studies to be made at redshifts beyond those where individual galaxies can be observed in HI with either existing or future radio telescopes.

\bigskip

\noindent
{\bf Acknowledgments.} The research was supported by the Australian
Research Council (JSBW).

\label{lastpage}
\end{document}